\documentclass[twocolumn,aps,english,prd,nofootinbib,preprintnumbers]{revtex4}
\usepackage{mathrsfs}
\usepackage{graphicx}
\usepackage{amsmath, amssymb}
\usepackage{babel}
\usepackage{color}
\usepackage{slashed}
\usepackage{hyperref}
\usepackage[T1]{fontenc}
\usepackage[export]{adjustbox}
\usepackage{multirow}

\begin{document}
%\preprint{ACFI-T18-12}
%\preprint{MITP/18-070}

\title{Towards {\it ab-initio} nuclear theory calculations of $\delta_\mathrm{C}$}

\author{Chien-Yeah Seng$^{1,2}$}
	\author{Mikhail Gorchtein$^{3,4}$}

	\affiliation{$^{1}$Facility for Rare Isotope Beams, Michigan State University, East Lansing, MI 48824, USA}
	\affiliation{$^{2}$Department of Physics, University of Washington,
		Seattle, WA 98195-1560, USA}
	\affiliation{$^{3}$Institut f\"ur Kernphysik, Johannes Gutenberg-Universit\"{a}t,\\
		J.J. Becher-Weg 45, 55128 Mainz, Germany}
	\affiliation{$^{4}$PRISMA Cluster of Excellence, Johannes Gutenberg-Universit\"{a}t, Mainz, Germany}

\date{\today}

\begin{abstract}

We propose a new theory framework to study the isospin-symmetry breaking correction $\delta_\text{C}$ in superallowed nuclear beta decays, crucial for the precise determination of $|V_{ud}|$. Based on a general assumptions of the isovector dominance in ISB interactions, we construct a set of functions $F_{T_z}$ which involve nuclear matrix elements of isovector monopole operators and the nuclear Green's function. Via the functions $F_{T_z}$, a connection of $\delta_\text{C}$ to measurable electroweak nuclear radii is established, providing an experimental gauge of the theory accuracy of $\delta_\text{C}$. We outline a strategy to perform ab-initio calculations of $F_{T_z}$ based on the Lanczos algorithm, and discuss its similarity with other nuclear-structure-dependent inputs in nuclear beta decays. 

\end{abstract}

\maketitle

\section{Introduction}

For many decades, superallowed beta decays of $J^p=0^+$, $T=1$ nuclei have provided the best measurement of the Cabibbo-Kobayashi-Maskawa (CKM) matrix element $V_{ud}$. The reason is of twofold: (1) At tree level only the vector charged weak current is involved, whose matrix element is exactly known assuming isospin symmetry, and (2) there are so far 23 measured superallowed transitions, with 15 among them whose $ft$-value precision is 0.23\% or better; the large sample size implies a large gain in statistics~\cite{Hardy:2020qwl}. 

This advantageous stance is now challenged by the free neutron decay. On the one hand, the latter benefits from recent improvements in the single-nucleon radiative correction theory~\cite{Seng:2018yzq,Seng:2018qru,Czarnecki:2019mwq,Seng:2020wjq,Hayen:2020cxh,Shiells:2020fqp,Cirigliano:2022hob} and measurements of the neutron lifetime $\tau_n$~\cite{Serebrov:2017bzo,Pattie:2017vsj,Ezhov:2014tna,UCNt:2021pcg} and the axial coupling constant $g_A$~\cite{UCNA:2017obv,Markisch:2018ndu,Beck:2019xye,Hassan:2020hrj}. On the other, recent analyses unveiled new sources of nuclear structure uncertainties in superallowed decays~\cite{Seng:2018qru,Gorchtein:2018fxl,Seng:2022inj}. In fact, taking the single best measurement of $\tau_n$ and $g_A$, one obtains (adopting the value of single-nucleon radiative correction quoted in Ref.\cite{Cirigliano:2022yyo}):
\begin{equation}
    |V_{ud}|_n=0.97413(43)~,
\end{equation}
which should be compared to the superallowed beta decay determination quoted in the same reference:
\begin{equation}
    |V_{ud}|_{0^+}=0.97367(32)~.
\end{equation}
One sees that the precision of $|V_{ud}|_{n}$ is indeed getting closer to $|V_{ud}|_{0^+}$ and, more importantly, a small discrepancy between the two values starts to emerge. This could add to the so-called Cabibbo angle anomaly~\cite{Crivellin:2020lzu,Crivellin:2021njn,Seng:2021gmh}, the mutual disagreement between different extractions of the Cabibbo angle $\theta_C$, which was also sharpened by new theory calculations in the $V_{us}$ sector~\cite{Seng:2019lxf,Seng:2020jtz,Ma:2021azh,Seng:2021boy,Seng:2021wcf,Seng:2021nar,Seng:2022wcw}.

Further improvements in the nuclear-structure-dependent Standard Model (SM) theory for superallowed nuclear decays are required for the latter to regain their lead. Process-specific quantities originating from measurements and theoretical corrections are usually lumped into the nucleus-independent $\mathcal{F}t$-value,
\begin{equation}
\mathcal{F}t\equiv ft(1+\delta_\text{R}')(1+\delta_\text{NS}-\delta_\text{C})\,.
\end{equation}
The nucleus-dependent $ft$-values are derived from experimental measurements of the decays' $Q$-values, branching ratios and halflives by absorbing Coulomb distortion effects in terms of the point-charge Fermi function \cite{Fermi:1934hr} and beyond (see \cite{Hayen:2017pwg} for a review). 
The ``outer radiative correction'' $\delta_\text{R}'$ accounts for QED effects beyond Coulomb distortions, and is well under control~\cite{Sirlin:1967zza,Sirlin:1987sy,Sirlin:1986cc,Towner:2007np}.
The remaining two corrections depend on nuclear structure in a non-trivial way: $\delta_\text{NS}$ represents the nuclear-structure correction to the single-nucleon $\gamma W$ box diagram, whereas $\delta_\text{C}$ represents the isospin-symmetry breaking (ISB) correction to the Fermi matrix element $M_F$. The recent inflation of the $|V_{ud}|_{0^+}$ theory uncertainty comes entirely from $\delta_\text{NS}$, where a previously missed correction from quasi-elastic nucleons was identified. A combination of the dispersive representation~\cite{Seng:2021syx,Seng:2022cnq} and ab-initio calculations of nuclear $\gamma W$-box diagrams may help reducing this uncertainty in the near future.

In this paper we concentrate on another important theory input, the ISB correction $\delta_\text{C}$. It measures the deviation of the full Fermi matrix element $M_F$ from its 
isospin-symmetric limit, $M_F^0=\sqrt{2}$ for $T=1$. So far its determination relies solely on model calculations, which is a classic problem in nuclear theory for more than 6 decades~\cite{MacDonald:1958zz}. Frequently quoted results include calculations based on the nuclear shell model with Woods-Saxon (WS) potential~\cite{Towner:2002rg,Towner:2007np,Hardy:2008gy,Xayavong:2017kim}, Hartree-Fock wave functions~\cite{Ormand:1989hm,Ormand:1995df}, density functional theory~\cite{Satula:2011br,Satula:2016hbs}, random-phase approximation~\cite{Liang:2009pf} and the isovector monopole resonance sum rule~\cite{Auerbach:2008ut}. Results from different methods are almost randomly-scattered and show no sign of convergence (see, e.g. Table I in Ref.\cite{Seng:2022epj}).

Interestingly enough, despite the tremendous model-dependence, 
the assigned theory uncertainty for $|V_{ud}|_{0^+}$ due to $\delta_\text{C}$ in a number of highly-cited global analysis turns out to be extremely small~\cite{Hardy:2014qxa,Hardy:2020qwl}. A criterion adopted in these analysis is the ability of the model calculation to align the $\mathcal{F}t$-values of different superallowed transitions, 
per request of the conserved vector current (CVC) hypothesis~\cite{Towner:2010bx}. This criterion effectively ruled out all but one calculation, namely the WS result, which they used in their subsequent analysis. However, this strategy is not without controversy: for example, one cannot rule out the possibility that the CVC hypothesis is invalidated by physics beyond the Standard Model (BSM), or that there is constant shift to all values of $\delta_\text{C}$. 
It has also been pointed out that the theory framework on which the WS calculation is based contains several inconsistencies, e.g. not using the correct isospin operator~\cite{Miller:2008my,Miller:2009cg,Condren:2022dji}, and correcting for these might lead to a substantial reduction of the $\delta_\text{C}$ values. %In this work we will also show that a number of the WS results seems to be disfavored by simple isospin argument. All in all, we think it is not unreasonable to conclude that the current theory uncertainty of $|V_{ud}|_{0^+}$ associated to $\delta_\text{C}$ had been largely underestimated.

A major limitation of existing calculations of $\delta_\text{C}$ is the absence of direct constraints from measurable ISB observables which can be used to quantify the theory uncertainties. The most precisely studied ISB observable in nuclear systems is the isobaric multiplet mass equation (IMME) that describes the mass splitting between isobaric analog states~\cite{MacDonald:1955zz,MacDonald:1955zza,MacDonald:1956zz,Weinberg:1959zzb}; it was used in a number of studies to either fix the model parameters~\cite{Towner:2007np} or as a preliminary test of the methodology's applicability~\cite{Martin:2021bud}. However, 
%it is well-known and we will also show later that, 
there is no overlap between the leading nuclear matrix elements that contribute to the IMME coefficients and to $\delta_\text{C}$, so the extent to which IMME constrains $\delta_\text{C}$ is not entirely clear. To overcome this limitation, we identified in Ref.~\cite{Seng:2022epj}
a set of ISB observables $\Delta M_{A,B}^{(1)}$ constructed from the electroweak nuclear radii across the isotriplet, which depend on the same nuclear matrix elements as $\delta_\text{C}$. Measurements of the former from atomic spectroscopy, beta decay recoil effects and fixed-target scattering experiments allow one to constrain the latter. To illustrate this idea, we adopted a simple isovector monopole dominance picture to derive a proportionality relation between $\Delta M_{A,B}^{(1)}$ and $\delta_\text{C}$. Despite being model-dependent, this simple picture offers a useful guidance for the precision target of future experiments. 

In this work we further explore the idea in Ref.\cite{Seng:2022epj} in a model-independent way. We construct a set of functions of an energy variable $\zeta$ $F_{T_z}(\zeta)$ ($T_z=-1,0,1$) that depend on the nuclear matrix elements common to $\delta_\text{C}$ and $\Delta M_{A,B}^{(1)}$. We show how the needed ISB observables can be derived from $F_{T_z}$ and its derivatives. Therefore, if a theory approach can reliably calculate $F_{T_z}$ as a function of $\zeta$, it simultaneously predicts $\Delta M_{A,B}^{(1)}$ and $\delta_\text{C}$ with a correlated degree of accuracy. A good agreement of the calculations with the experimental measurements for the former will imply the reliability of the theory prediction for the latter. In this sense, the approach advocated here directly constrains $\delta_\text{C}$ and its uncertainty by the experiment. 

The content of this work is arranged as follows. In Section~\ref{sec:PT} we derive the leading perturbative expression of $\delta_\text{C}$ and argue that existing model calculations may contain large systematic uncertainties. In  Section~\ref{sec:radii} we review the central idea in Ref.\cite{Seng:2022epj}, namely the construction of the two ISB observables $\Delta M_{A,B}^{(1)}$ from the measurable electroweak nuclear radii. In Section~\ref{sec:Fz} we define the functions $F_{T_z}(\zeta)$ and demonstrate their connection to $\delta_\text{C}$ and $\Delta M_{A,B}^{(1)}$. In Section~\ref{sec:strategy} we discuss possible strategies to compute $F_{T_z}(\zeta)$ as a function of $\zeta$, which simultaneously predicts $\Delta M_{A,B}^{(1)}$ and $\delta_\text{C}$. In Section~\ref{sec:conclusion} we draw our conclusions.

\section{\label{sec:PT}ISB in perturbation theory}

To discuss the perturbative expression of ISB observables, we split the full Hamiltonian as $H=H_0+V$, where $H_0$ is the unperturbed, isospin-conserving part and $V$ is the ISB perturbation term. We label the eigenstates of $H_0$ as $|a;T,T_z\rangle$ (with unperturbed energy $E_{a,T}$), where $T,T_z$ are the isospin quantum numbers, and $a$ represents all other quantum numbers unrelated to isospin. In particular, the ground state isotriplet that undergoes superallowed beta decay transitions is labelled as $|g;1,T_z\rangle$.

The most commonly studied ISB observable is IMME,
\begin{equation}
    E(a,T,T_z)=\mathtt{a}(a,T)+\mathtt{b}(a,T)T_z+\mathtt{c}(a,T)T_z^2~,\label{eq:IMME}
\end{equation}
which takes its form based on the fact that any two-nucleon interaction can at most be isotensor, i.e. we can write $V=V^{(1)}+V^{(2)}$, where the superscript denotes the isospin.
The coefficients $\mathtt{b}$ and $\mathtt{c}$ characterize the strength of ISB effects. To first order in perturbation theory, they are related to the diagonal matrix element of $V$:
\begin{equation}
    \mathtt{b}\sim \langle a;T,T_z|V^{(1)}|a;T,T_z\rangle~,~\mathtt{c}\sim \langle a;T,T_z|V^{(2)}|a;T,T_z\rangle~.
\end{equation}
Experimental measurements show in general $|\mathtt{b}|\gg |\mathtt{c}|$ which indicates the dominance of isovector ISB effects. For instance, in $J^P=0^+$, $T=1$ isomultiplets, one observes that the ratio $|\mathtt{b}/\mathtt{c}|\geq 15$ for $A\geq 26$, and increases with increasing $A$~\cite{Lam:2013bhc}.

\begin{table}
\begin{centering}
\begin{tabular}{|c|c|c|c|c|c|c|}
\hline 
\multirow{2}{*}{$A$} & \multicolumn{3}{c|}{WS} & \multicolumn{3}{c|}{RPA}\tabularnewline
\cline{2-7} 
 & $\delta_{\text{C}}^{-1}(\%)$ & $\delta_{\text{C}}^{0}(\%)$ & $\Delta_{C}$ & $\delta_{\text{C}}^{-1}(\%)$ & $\delta_{\text{C}}^{0}(\%)$ & $\Delta_{C}$\tabularnewline
\hline 
\hline 
26 & 0.435(27) & 0.310(18) & 0.34(8) & 0.176 & 0.139 & 0.23\tabularnewline
\hline 
34 & 0.659(40) & 0.613(49) & 0.07(10) & 0.268 & 0.234 & 0.14\tabularnewline
\hline 
38 & 0.745(47) & 0.628(54) & 0.17(11) & 0.313 & 0.278 & 0.12\tabularnewline
\hline 
42 & 0.960(63) & 0.690(46) & 0.32(9) & 0.384 & 0.333 & 0.14\tabularnewline
\hline 
46 & 0.760(87) & 0.620(63) & 0.20(15) & / & / & /\tabularnewline
\hline 
50 & 0.660(49) & 0.660(32) & 0.00(0) & / & / & /\tabularnewline
\hline 
54 & 0.790(67) & 0.770(67) & 0.03(4) & / & 0.319 & /\tabularnewline
\hline 
\end{tabular}
\par\end{centering}
\caption{\label{tab:deltaCab}$\delta_\text{C}^{-1,0}$ and $\Delta_{C}$ computed with WS and RPA.}

\end{table}

On the other hand,  $\delta_\text{C}$ depends on a completely different set of nuclear matrix elements than the IMME coefficients $\mathtt{b}$ and $\mathtt{c}$. To see this, we start with the exact formalism by Miller and Schwenk~\cite{Miller:2008my}, and label the eigenstates of $H$ and $H_0$ temporarily as $|n\rangle$ and $|n)$ respectively. The full Fermi matrix element for a superallowed transition $i\rightarrow f$ is given by 
\begin{equation}
M_F=\langle f|\tau_+|i\rangle~,\label{eq:MF}
\end{equation}
with $\tau_+$ the isospin raising operator. Similarly, the isospin-limit Fermi matrix element is $M_F^0=(f|\tau_+|i)$. The Wigner-Brillouin perturbation theory implies,
\begin{equation}
    |n\rangle=\sqrt{\mathcal{Z}_n}\left[|n)+\frac{1}{E_n-\Lambda_n H\Lambda_n}\Lambda_n V|n)\right]~,\label{eq:WigBri}
\end{equation}
where $E_n$ is the energy of the full state $|n\rangle$, $\Lambda_n=1-|n)(n|$ projects out the unperturbed state $|n)$, and 
\begin{equation}
    \mathcal{Z}_n=\left[1+(n|V\Lambda_n\left(\frac{1}{E_n-\Lambda_n H\Lambda_n}\right)^2\Lambda_n V|n)\right]^{-1}
\end{equation}
is a normalization factor to ensure $\langle n|n\rangle=1$. Substituting Eq.\eqref{eq:WigBri} into Eq.\eqref{eq:MF} gives:
\begin{eqnarray}
M_F&=&\sqrt{\mathcal{Z}_i\mathcal{Z}_f}\left[M_F^0+(f|V\Lambda_f\frac{1}{E_f-\Lambda_fH\Lambda_f}\right.\nonumber\\
&&\left.\times \tau_+\frac{1}{E_i-\Lambda_iH\Lambda_i}\Lambda_iV|i)\right]~,\label{eq:fullMFM0}
\end{eqnarray}
which is the central result of Ref.\cite{Miller:2008my}. It is clear from the expression above that the deviation between $M_F$ and $M_F^0$ starts at $\mathcal{O}(V^2)$. 
Concentrating on the $\mathcal{O}(V^2)$ corrections in Eq.\eqref{eq:fullMFM0} and using the definition $|M_F|^2=|M_F^0|^2(1-\delta_\text{C})$, we get,
\begin{eqnarray}
&&\delta_\text{C}=\langle g;1,T_{zi}|V\Lambda_i\left(\frac{1}{E_{g,1}-\Lambda_i H_0\Lambda_i}\right)^2\Lambda_iV|g;1,T_{zi}\rangle\nonumber\\
&&+\langle g;1,T_{zf}|V\Lambda_f\left(\frac{1}{E_{g,1}-\Lambda_f H_0\Lambda_f}\right)^2\Lambda_fV|g;1,T_{zf}\rangle\nonumber\\
&&-\frac{2}{M_F^0}\langle g;1,T_{zf}|V\Lambda_f\frac{1}{E_{g,1}-\Lambda_fH\Lambda_f}\tau_+\nonumber\\
&&\times\frac{1}{E_{g,1}-\Lambda_iH\Lambda_i}\Lambda_iV|g;1,T_{zi}\rangle+\mathcal{O}(V^3)~.\label{eq:deltaCpert}
\end{eqnarray}
We observe that due to the presence of the projection operators the leading expression of $\delta_\text{C}$ contains no diagonal nuclear matrix element of the form $\langle g;1,T_z|V|g;1,T_z\rangle$, so it is orthogonal to the leading expressions of the IMME coefficients $\{\mathtt{b},\mathtt{c}\}$. Subsequently, the ability of a model calculation to reproduce the IMME coefficients accurately does not guarantee its ability to determine $\delta_\text{C}$ with the same accuracy. 

To proceed further we must invoke some general properties of the ISB interaction $V$. We will assume that $V$ is predominantly isovector, i.e. $V\approx V^{(1)}$. The IMME coefficients suggest that, for a $\sim 10\%$ precision goal this is a good assumption for $A\geq 26$.  
With this, we insert a complete set of intermediate nuclear states $\{|a;T,T_z\rangle\}$ to each term in Eq.\eqref{eq:deltaCpert} and apply the Wigner-Eckart theorem,
\begin{equation}
    \langle a;T,T_z|V|g;1,T_z'\rangle=C_{1,T_z';1,0}^{1,1;T,T_z}\langle a;T||V||g;1\rangle~,
\end{equation}
with $C$s the Clebsch-Gordan coefficients. It recasts $\delta_\text{C}$ in terms of the reduced matrix element $\langle a;T||V||g;1\rangle$. Since $V$ is an isovector, the intermediate states can only have $T=0,1,2$; also, the $a=g$, $T=1$ intermediate states are excluded by the projection operators. With these we obtain:
\begin{eqnarray}
\delta_\text{C}&=&\frac{1}{3}\sum_a\frac{|\langle a;0||V||g;1\rangle|^2}{(E_{a,0}-E_{g,1})^2}+\frac{1}{2}\sum_{a\neq g}\frac{|\langle a;1||V||g;1\rangle|^2}{(E_{a,1}-E_{g,1})^2}\nonumber\\
&&-\frac{5}{6}\sum_a\frac{|\langle a;2||V||g;1\rangle|^2}{(E_{a,2}-E_{g,1})^2}+\mathcal{O}(V^3)~.\label{eq:deltaCV2}
\end{eqnarray}

Within an isotriplet there are two superallowed transitions: $(T_{zi}=-1)\rightarrow (T_{zf}=0)$ and $(T_{zi}=0)\rightarrow (T_{zf}=+1)$. 
It turns out that Eq.\eqref{eq:deltaCV2} applies to both transitions, which means that $\delta_\text{C}$ for the superallowed beta decays within the same isotripet are identical up to $\mathcal{O}(V^2)$ assuming the dominance of isotripet ISB interaction. This conclusion is model-independent as it straightforwardly follows from the Wigner-Eckart theorem, and serves as a useful consistency check of existing calculations. Interestingly enough, this simple conclusion has never been discussed in literature. As an example, we quote in Table~\ref{tab:deltaCab} the WS~\cite{Hardy:2020qwl} and RPA (with PKO1 parameterization)~\cite{Liang:2009pf} calculation of $\delta_\text{C}$ for the $T_{zi}=-1$ and $T_{zi}=0$ transitions (which we denote as $\delta_{C}^{-1}$ and $\delta_\text{C}^0$ respectively) within the same isotriplet, and define their relative difference: $\Delta_{C}\equiv |2(\delta_\text{C}^{-1}-\delta_\text{C}^0)/(\delta_\text{C}^{-1}+\delta_\text{C}^0)|$. We find that some of their results give $\Delta_{C}$ as large as 20\% or more for $A\geq 26$. We conclude that even the most widely-adopted model calculation of $\delta_\text{C}$ is not free from potentially large systematic errors. 

\section{\label{sec:radii}Electroweak nuclear radii probe ISB effects}

The first step towards a systematic reevaluation of ISB corrections to superallowed beta decays is to identify new experimental observables that are able to directly constrain $\delta_\text{C}$. This idea is pioneered in Ref.\cite{Seng:2022epj}, and we briefly review it below.    
A key object throughout the discussion is the isovector monopole operator defined as $\vec{M}^{(1)}=\sum_i r_i^2\vec{\hat{T}}(i)$, where $\vec{\hat{T}}$ is the isospin operator and $i$ labels the nucleons in the nucleus. 
Rank-1 irreducible tensors in the isospin space can be formed as: $M_0^{(1)}=M_z^{(1)}$, $M_{\pm 1}^{(1)}=\mp (M_x^{(1)}\pm i M_y^{(1)})/\sqrt{2}$. For convenience, we may also define a corresponding isoscalar monopole operator as $M^{(0)}=\sum_i r_i^2$. 

In Ref.\cite{Seng:2022epj} we defined two ISB-sensitive combinations of experimental observables. The first one reads
\begin{equation}
\Delta M_A^{(1)}\equiv \langle f|M_{+1}^{(1)}|i\rangle+\langle f|M_0^{(1)}|f\rangle~.
\end{equation}
The first term on the right hand side comes from the measurement of the $t$-dependence of the $(T_{zi}=0)\rightarrow (T_{zf}=+1)$ superallowed beta decay form factor,
\begin{equation}
\bar{f}_+(t)=1-\frac{t}{6}\langle f|M_{+1}^{(1)}|i\rangle +\mathcal{O}(t^2)~,
\end{equation} 
which corresponds to the charged weak radius. The second term combines the proton and neutron distribution radii of the $T_z=+1$ daughter nucleus,
\begin{equation}
\langle f|M_0^{(1)}|f\rangle =\frac{N_f}{2}R_{n,f}^2-\frac{Z_f}{2}R_{p,f}^2~.
\end{equation}
Above, the root mean square (RMS) distribution radius of a nucleon in a nucleus $\phi$ is defined as:
\begin{equation}
R_{p/n,\phi}=\sqrt{\frac{1}{X}\langle \phi|\sum_{i=1}^Ar_i^2\left(\frac{1}{2}\pm\hat{T}_z(i)\right)|\phi\rangle}~,
\end{equation}
with $-$ for the proton and $+$ for the neutron, and $X=Z_\phi$ or $N_\phi$. These radii can be measured through fixed-target scattering experiments. 

In the meantime, we recall that the nuclear charge radius, largely given by $R_p$, is measurable via atomic spectroscopy for both stable and unstable nuclei. With this, one may construct another experimental observable by combining the charge radii across the isotriplet, 
\begin{equation}
\Delta M_B^{(1)}\equiv \frac{1}{2}\left(Z_1 R_{p,1}^2+Z_{-1}R_{p,-1}^2\right)-Z_0 R_{p,0}^2\label{eq:DeltaMB1}
\end{equation}
where the subscript $-1,0,1$ denotes $T_z$ of the nucleus.

It is easy to see using Wigner-Eckart theorem that both $\Delta M_{A,B}^{(1)}$ vanish identically in the isospin limit: replacing the external states by isospin eigenstates, we get
\begin{eqnarray}
\Delta M_A^{(1)}&\rightarrow&\langle g;;1,1|M_{+1}^{(1)}|g;1,0\rangle +\langle g;1,1|M_0^{(1)}|g;1,1\rangle\nonumber\\
&=&0\nonumber\\
\Delta M_B^{(1)}&\rightarrow &\sum_{T_z=\pm 1}\langle g;1,T_z|\frac{1}{4}M^{(0)}-\frac{1}{2}M_0^{(1)}|g;1,T_z\rangle\nonumber\\
&&-\langle g;1,0|\frac{1}{2}M^{(0)}-M_0^{(1)}|g;1,0\rangle\nonumber\\
&=&0~.
\end{eqnarray}
This qualifies both observables as clean probes of ISB effects. 
Their leading non-zero expression arises by expanding the external states to $\mathcal{O}(V)$ following Eq.\eqref{eq:WigBri}. Assuming the isovector dominance in $V$, as discussed in Sec.\ref{sec:PT}, a straightforward derivation gives
\begin{eqnarray}
&&\Delta M^{(1)}_A = -\frac{1}{3}\sum_a \frac{\langle a;0||M^{(1)}||g;1\rangle^{*}\langle a;0||V||g;1\rangle}{E_{a,0}-E_{g,1}}\nonumber\\
&& -\frac{1}{2}\sum_{a\neq g}\frac{\langle a;1||M^{(1)}||g;1\rangle^{*}\langle a;1||V||g;1\rangle}{E_{a,1}-E_{g,1}}\nonumber\\
&&-\frac{1}{6}\sum_{a}\frac{\langle a;2||M^{(1)}||g;1\rangle^{*}\langle a;2||V||g;1\rangle}{E_{a,2}-E_{g,1}}\nonumber\\
&&-\sum_{a}\frac{\langle a;2||V||g;1\rangle^{*}\langle a;2||M^{(1)}||g;1\rangle}{E_{a,2}-E_{g,1}} +\mathcal{O}(V^2)\label{eq:DeltaM1Apert}
\end{eqnarray}
and 
\begin{eqnarray}
&&\Delta M^{(1)}_B = \mathfrak{Re}\left\{-\frac{2}{3}\sum_a \frac{\langle a;0||M^{(1)}||g;1\rangle^{*}\langle a;0||V||g;1\rangle}{E_{a,0}-E_{g,1}}\right.\nonumber\\
&& +\sum_{a\neq g}\frac{\langle a;1||M^{(1)}||g;1\rangle^{*}\langle a;1||V||g;1\rangle}{E_{a,1}-E_{g,1}}\label{eq:DeltaM1Bpert}\\
&&\left.-\frac{1}{3}\sum_{a}\frac{\langle a;2||M^{(1)}||g;1\rangle^{*}\langle a;2||V||g;1\rangle}{E_{a,2}-E_{g,1}}\right\} +\mathcal{O}(V^2)\,,\nonumber
\end{eqnarray}
respectively. The reduced matrix elements of $\vec{M}^{(1)}$ are defined as
\begin{equation}
\langle a;T'',T_z''|M^{(1)}_{T_z}|g;1,T_z'\rangle=C_{1,T_z';1,T_z}^{1,1;T'',T_z''}\langle a;T''||M^{(1)}||g;1\rangle~.
\end{equation}
It is easy to check that the definition of $\Delta M_B^{(1)}$ in Eq.\eqref{eq:DeltaMB1} ensures the absence of terms $\sim M^{(0)}\otimes V$ at $\mathcal{O}(V)$. 

\section{\label{sec:Fz}Universal functions connecting all ISB observables}

The dominant source of ISB is the Coulomb repulsion between protons, with its prevailing part coming from the one-body potential of a uniformly charged sphere of radius $R_C$,
\begin{equation}
V_C\approx -\frac{Ze^2}{4\pi R_C^3}\sum_{i=1}^A\left(\frac{1}{2}r_i^2-\frac{3}{2}R_C^2\right)\left(\frac{1}{2}-\hat{T}_z(i)\right)~.\label{eq:VCinside}
\end{equation}
For the isotriplets of interest we may take $Z\approx A/2$, and $R_C$ is related to the point-proton radius of the respective nucleus as $R_C^2=(5/3)R_p^2$.
Notice that the potential above assumes that all nucleons reside at $r_i<R_C$. In reality, there are (small) corrections due to the non-zero nucleon wave functions at $r_i>R_C$ where the potential behaves as $1/r_i$. This residual effect could be estimated within nuclear models and included as a part of the systematic uncertainties in the theory analysis.

The ISB part of $V_C$ involves only the $\hat{T}_z(i)$ in the second round bracket, and is purely isovector. Furthermore, as far as the off-diagonal matrix elements are concerned, the $R_C$ term in the first bracket does not contribute. One may therefore make the connection,
\begin{equation}
V\leftrightarrow (Ze^2/8\pi R_C^3)M_0^{(1)}~.\label{eq:VCM01ID}
\end{equation}
With this, both $\delta M_{A,B}^{(1)}$ and $\delta_\text{C}$ share the same set of reduced matrix elements of the form $|\langle a;T||M^{(1)}||g;1\rangle|^2$, which is the central result of Ref.~\cite{Seng:2022epj}. The main difference is that $\Delta M_{A,B}^{(1)}$ contain only one energy denominator because they arise from a first-order perturbation, while $\delta_\text{C}$ starts from second order and contains two energy denominators. We define the following {\it generating function}
\begin{equation}
    \bar{\Gamma}_T(\zeta)\equiv\sum_{a\neq g}\frac{|\langle a;T||M^{(1)}||g;1\rangle|^2}{E_{a,T}-\zeta}~,~T=0,1,2\label{eq:Gammabar}
\end{equation}
with $\zeta$ an energy variable. 
The value of $\bar{\Gamma}_T(\zeta)$ at $\zeta=E_{g,1}$ is directly related to $\Delta M_{A,B}^{(1)}$, while its derivative at $\zeta=E_{g,1}$ is directly related to $\delta_\text{C}$. 

To directly access the reduced matrix elements in Eq.\eqref{eq:Gammabar} through nuclear theory calculations, we define a set of nuclear matrix elements for $T_z=-1,0,1$ which will be the key objects of theory studies,
\begin{eqnarray}
    F_{T_z}(\zeta)&\equiv& \langle g;1,T_z|(M_{-1}^{(1)})^\dagger G(\zeta) M_{-1}^{(1)}|g;1,T_z\rangle\nonumber\\ &-&\frac{|\langle g;1,T_z-1|M_{-1}^{(1)}|g;1,T_z\rangle|^2}{\zeta-E_{g,1}}~,\label{eq:FTz}
\end{eqnarray}
with $G(\zeta)=1/(\zeta-H_0)$ the nuclear Green's function. The second term on the right hand side subtracts out the $a=g$, $T=1$ intermediate state contribution and exists only for $T_z=1,0$ but not for $T_z=-1$. Inserting a complete set of nuclear states to the first term, we get
\begin{eqnarray}
F_1(\zeta)&=&-\frac{1}{3}\bar{\Gamma}_0(\zeta)-\frac{1}{2}\bar{\Gamma}_1(\zeta)-\frac{1}{6}\bar{\Gamma}_2(\zeta)\nonumber\\
F_0(\zeta)&=&-\frac{1}{2}\bar{\Gamma}_1(\zeta)-\frac{1}{2}\bar{\Gamma}_2(\zeta)\nonumber\\
F_{-1}(\zeta)&=&-\bar{\Gamma}_2(\zeta)~.
\end{eqnarray}
These three matrix elements can be solved for all three $\bar{\Gamma}_T(\zeta)$. To connect to our ISB observables we expand the functions $F_{T_z}$ around $\zeta=E_{g,1}$,
\begin{equation}
    F_{T_z}(\zeta)=\alpha_{T_z}+\beta_{T_z}(\zeta-E_{g,1})+\mathcal{O}((\zeta-E_{g,1})^2)~,
\end{equation}
where $\alpha_{T_z}$, $\beta_{T_z}$ are constant expansion coefficients.
With these, we obtain
\begin{eqnarray}
    \Delta M_A^{(1)}&=&\frac{Ze^2}{8\pi R_C^3}\left\{\alpha_1+\alpha_{-1}\right\}\nonumber\\
    \Delta M_B^{(1)}&=&\frac{Ze^2}{8\pi R_C^3}\left\{2\alpha_1-4\alpha_0+2\alpha_{-1}\right\}~,
\end{eqnarray}
and 
\begin{equation}
    \delta_\text{C}=-\left(\frac{Ze^2}{8\pi R_C^3}\right)^2\left\{\beta_1-\beta_{-1}\right\}~.
\end{equation}

The interconnection between $\delta_\text{C}$ and the electroweak nuclear radii through the universal function $F_{T_z}(\zeta)$ is largely model-independent, as long as the identification of Eq.\eqref{eq:VCM01ID} holds, and may be viewed as a kind of sum rule. A theory calculation of $F_{T_z}(\zeta)$ that simultaneously gives all the expansion coefficients, will be able to predict $\delta_\text{C}$, and at the same time receive direct experimental constraints from $\Delta M_{A,B}^{(1)}$. These experimental constraints allow us to quantify the theory uncertainties in $\delta_\text{C}$. 

\section{Limitations of the conventional shell model}

To illustrate the idea we propose and investigate the main physics that enters $\delta_\text{C}$ by computing the generating function $F_{T_z}(\zeta)$ explicitly using conventional shell model. Through this process we are also able to check the validity of some assumptions we made in arriving Eq.\eqref{eq:FTz}. We consider the $A=38$ isotriplet as an example. 

\subsection{Checking the approximation to the Coulomb potential}

The rigorous expression of a Coulomb potential felt by a point-like proton from a nucleus of charge $Z$ is ($0<r<\infty$):
\begin{eqnarray}
V_C(r)&=&-\frac{Ze^2}{8\pi R_C^3}(r^2-3R_C^2)\Theta(R_C-r)\nonumber\\
&&+\frac{Ze^2}{4\pi r}\Theta(r-R_C)~.
\end{eqnarray}
The approximation in Eq.\eqref{eq:VCinside} then corresponds to replacing the equation above by ($0<r<\infty$):
\begin{eqnarray}
V_C^\text{appr}(r)&\equiv&-\frac{Ze^2}{8\pi R_C^3}(r^2-3R_C^2)~.
\end{eqnarray}
We adopt the empirical formula for the sphere radius:
\begin{equation}
R_C=\sqrt{\frac{5}{3}}\langle r_p^2\rangle^{1/2}\approx \sqrt{\frac{5}{3}}\times 1.1A^{1/3}~\text{fm}~.
\end{equation}

In the shell model, we describe single-nucleon states by the harmonic oscillator wavefunction: $\psi_{n\ell m}(\vec{x})=R_{n\ell}(r)Y_{\ell m}(\hat{x})$. The radial wavefunction reads:
\begin{eqnarray}
R_{n\ell}(r)&=&\mathcal{N}r^\ell e^{-\nu r^2}L_{n-1}^{\ell+\frac{1}{2}}(2\nu r^2)\nonumber\\
\mathcal{N}&=&\left[\left(\frac{2\nu^3}{\pi}\right)^{1/2}\frac{2^{n+2\ell+2}(n-1)!!\nu^\ell}{(2n+2\ell-1)!!}\right]^{1/2}~,
\end{eqnarray}
where $\nu=m\omega/2$ with $m$ the nucleon mass, and the harmonic oscillator frequency:
\begin{equation}
\omega\approx 41A^{-1/3}~\text{MeV}~.
\end{equation}
with these, we can compute the single-particle expectation value of the Coulomb potential $V_C$ and the approximated expression $V_C^\text{appr}$:
\begin{eqnarray}
\langle V_C\rangle_{n\ell}&=&-\frac{Ze	2}{8\pi R_C^3}\int_0^{R_C}(r^2-3R_C^2)R_{n\ell}^2r^2dr\nonumber\\
&&+\frac{Ze^2}{4\pi}\int_{R_C}^\infty \frac{1}{r}R_{n\ell}^2r^2dr\nonumber\\
\langle V_C^\text{appr}\rangle_{n\ell}&=&-\frac{Ze	2}{8\pi R_C^3}\int_0^{\infty}(r^2-3R_C^2)R_{n\ell}^2r^2dr~,
\end{eqnarray}
from which we can define a relative error:
\begin{equation}
\Delta_{n\ell}\equiv
%\left(
\frac{\langle V_C^\text{appr}\rangle_{n\ell}}{\langle V_C\rangle_{n\ell}}-1~.
%\right)\times 100\%~.
\end{equation}

The nucleons in the $A=38$ isotriplet occupy the orbitals $1s$, $1p$, $1d$ and $2s$. For these orbitals we find:
\begin{eqnarray}
&&\Delta_{1s}\approx -0.004\%~,~\Delta_{1p}\approx -0.03\%~,\nonumber\\
&&~\Delta_{1d}\approx -0.1\%~,~\Delta_{2s}\approx -0.2\%~,
\end{eqnarray}
much below our aimed precision goal for $\delta_\text{C}$ ($\sim 10\%$). We conclude that within the shell model framework the approximation of $V_C$ in Eq.\eqref{eq:VCinside} is valid. A similar check can be done for other non-shell-model approaches. 

\subsection{Computing the generating function}

Now we proceed to compute the generating function. Recall the nuclear Green's function
\begin{equation}
G(\zeta)=\frac{1}{\zeta-H_0}=\sum_X\frac{|X\rangle \langle X|}{\zeta-E_X}
\end{equation}
where $\{X\}$ are all possible intermediate states. Conventional shell model provides the wavefunctions of the single-nucleus ground and excited states with definite $J^P$ and isospin; restricting ourselves to these intermediate states, we may directly compute the isospin generating functions $\bar{\Gamma}_I$ in Eq.\eqref{eq:Gammabar} without going through $F_{T_z}$. For instance, for the $A=38$ system they can all be expressed in terms of the matrix element of $M_{-1}^{(1)}$ between the $^{38}$Ar ground state and all the $^{38}$K$(0^+)$ excited states:
\begin{eqnarray}
\bar{\Gamma}_0(\zeta)&=&-3\sum_a\frac{|\langle {}^{38}\text{K}(0^+,a,0)|M_{-1}^{(1)}|{}^{38}\text{Ar}(g)\rangle|^2}{\zeta-E({}^{38}\text{K}(0^+,a,0))}\nonumber\\
\bar{\Gamma}_1(\zeta)&=&-2\sum_{a\neq g}\frac{|\langle {}^{38}\text{K}(0^+,a,1)|M_{-1}^{(1)}|{}^{38}\text{Ar}(g)\rangle|^2}{\zeta-E({}^{38}\text{K}(0^+,a,0))}\nonumber\\
\bar{\Gamma}_2(\zeta)&=&-6\sum_a\frac{|\langle {}^{38}\text{K}(0^+,a,2)|M_{-1}^{(1)}|{}^{38}\text{Ar}(g)\rangle|^2}{\zeta-E({}^{38}\text{K}(0^+,a,0))}~.\label{eq:GammaIshell}
\end{eqnarray}

Next we discuss the the isovector monopole operator $M_{-1}^{(1)}$. In shell model we adopt the following second-quantized form:
\begin{eqnarray}
M_{-1}^{(1)}&=&\frac{1}{\sqrt{2}}\sum_{\alpha}\langle\alpha|r^2|\alpha\rangle b_\alpha^\dagger a_\alpha\nonumber\\
&=&\sum_{n\ell}\langle r^2\rangle_{n\ell}\sum_j\sqrt{\frac{2j+1}{2}}[b_{n\ell j}^\dagger \tilde{a}_{n\ell j}]_0^{(0)}~,\label{eq:Msecondquan}
\end{eqnarray}
where $a_\alpha$ annihilates a neutron at the single-nucleon orbital $\alpha=(n,\ell,j,m_j)$ within a chosen model space, while $b_\alpha^\dagger$ creates a proton; in the second line we recast the product of operators in terms of rank-0 tensors, with $a_{j,m_j}\equiv (-1)^{j-m_j}\tilde{a}_{j,-m_j}$. The single-nucleon matrix element $\langle r^2\rangle_{n\ell}$ can be computed using harmonic oscillator wavefunctions, and in fact the outcome is analytically known:
\begin{equation}
\langle r^2\rangle_{n\ell}=\frac{2n+\ell-\frac{1}{2}}{m\omega}=\frac{N+\frac{3}{2}}{m\omega}~,\label{eq:r2nl}
\end{equation}
where $N=2n+\ell-2$ is the principal quantum number. The expression above immediately tells us that, to get a possibly non-zero generating function, one cannot choose the model space to include only orbitals with the same $N$ (e.g. the $sd$-shell), because otherwise the matrix element $\langle \alpha|r^2|\alpha\rangle$ in Eq.\eqref{eq:Msecondquan} would become a constant and could be factorized out from the sum, which would make $M_{-1}^{(1)}\propto \tau_-$, where
\begin{equation}
\tau_-=\sum_{\alpha}b_\alpha^\dagger a_\alpha=\sum_{n\ell j}\sqrt{2j+1}[b_{n\ell j}^\dagger \tilde{a}_{n\ell j}]_0^{(0)}\label{eq:taum}
\end{equation}
is the isospin-lowering operator. Since $\tau_-$ only connects $|{}^{38}\text{Ar}(g)\rangle$ to $|{}^{38}\text{K}(0^+,g,1)\rangle$ which is excluded in Eq.\eqref{eq:GammaIshell}, such a choice of model space would result in vanishing generating functions. 

\begin{table}
	\begin{centering}
		\begin{tabular}{|c|c|c|}
			\hline 
			State label & $T$ & Excitation energy (MeV)\tabularnewline
			\hline 
			\hline 
			$^{38}\text{K}(0^{+},g,1)$ & 1 & 0.000\tabularnewline
			\hline 
			$^{38}\text{K}(0^{+},1,1)$ & 1 & 6.029\tabularnewline
			\hline 
			$^{38}\text{K}(0^{+},2,1)$ & 1 & 16.464\tabularnewline
			\hline 
		\end{tabular}
		\par\end{centering}
	\caption{\label{tab:Klevels}Shell model calculation of the ${}^{38}$K$(0^+)$ energy levels.}
	
\end{table}

Below we describe a sample calculation. We choose the model space $spsdpf$, which means all nucleons in the $A=38$ isotriplet are taken as valence quarks; in accordance to this model space, we choose the ``FSU interaction'' outlined in Refs.\cite{Lubna:2019twe,Lubna:2020tae}. This interaction is implemented to the latest version of the NuShellX code~\cite{Brown:2014bhl} which is able to perform shell model calculations with large dimensions. Using the code, we first study the energy levels of $^{38}$K$(0^+)$. The outcome is summarized in Table~\ref{tab:Klevels}, where we find that only $T=1$ states are predicted; they can only contribute to $\bar{\Gamma}_1(\zeta)$. 

\begin{table}
	\begin{centering}
		\begin{tabular}{|c|c|c|c|}
			\hline 
			& $\mathbf{\underline{a}}(1d_{5/2})$ & $\mathbf{\underline{a}}(1d_{3/2})$ & $\mathbf{\underline{a}}(2s_{1/2})$\tabularnewline
			\hline 
			\hline 
			$^{38}\text{Ar}(g)\rightarrow$$^{38}\text{K}(0^{+},g,1)$ & 0.01554 & 0.66856 & 0.02760\tabularnewline
			\hline 
			$^{38}\text{Ar}(g)\rightarrow$$^{38}\text{K}(0^{+},1,1)$ & -0.00739 & 0.12422 & -0.16289\tabularnewline
			\hline 
			$^{38}\text{Ar}(g)\rightarrow$$^{38}\text{K}(0^{+},2,1)$ & -0.09315 & 0.10169 & 0.01753\tabularnewline
			\hline 
		\end{tabular}
		\par\end{centering}
	\caption{\label{tab:OBDME}The non-vanishing OBDME for the ${}^{38}\text{Ar}(g)\rightarrow{}^{38}\text{K}(0^+)$ transition.}
	
\end{table}

Next, using the same model space, interaction and code, we compute the one-body density matrix element (OBDME) for the ${}^{38}\text{Ar}(g)\rightarrow{}^{38}\text{K}(0^+)$ transition:
\begin{equation}
\mathbf{\underline{a}}(n\ell j)\equiv \langle f|[b_{n\ell j}^\dagger \tilde{a}_{n\ell j}]_0^{(0)}|i\rangle~.
\end{equation}
The non-vanishing matrix elements are summarized in Table~\ref{tab:OBDME}, and let us understand what these numbers mean. First, we may compute the Fermi matrix element for the ${}^{38}\text{Ar}(g)\rightarrow{}^{38}\text{K}(0^+,g,1)$ transition:
\begin{equation}
M_F=\langle {}^{38}\text{K}(0^+,g,1)|\tau_-|{}^{38}\text{Ar}(g)\rangle
\end{equation}
using the second-quantized form of $\tau_-$ (Eq.\eqref{eq:taum}) and the numbers in the first row of Table~\ref{tab:OBDME}. The numerical result is close to $\sqrt{2}$, which is exactly what we expected (note that the FSU interaction is isospin-symmetric). 

In the meantime, plugging the numbers in the second and third row from Table~\ref{tab:OBDME} into Eq.\eqref{eq:Msecondquan}, we find that the transition matrix element of $M_{-1}^{(1)}$ from ${}^{38}\text{Ar}(g)$ to the ${}^{38}\text{K}(0^+)$ excited states are zero, which implies a vanishing generating function $\bar{\Gamma}_1(\zeta)$. This is easy to understand by realizing that, despite the large model space that we chose, the non-vanishing OBDME in Table~\ref{tab:OBDME} only involves orbitals in the $sd$-shell, so the $M_{-1}^{(1)}$ matrix element vanishes following our discussion after Eq.\eqref{eq:r2nl}. We have also tried different model space and interactions (e.g. $sdpf$ space with the SDPF-MU interaction~\cite{Utsuno:2012qf}, and $psd$ space with the interaction in Ref.\cite{Utsuno:2011zz}), but still obtain the same vanishing result. 

The working example above indicates that the conventional shell model, which only takes into account the lowest nuclear excitations, falls short to capture the main physics that induces a non-zero $\delta_\text{C}$. Some reasonable speculations are: (1) The sum of contributions from all higher nuclear excitations could return a non-suppressed effect, or (2) The continuum intermediate states (e.g. nucleus + nucleon) could play an important role. This also resonates with a previous observation that the direct computation of $\delta_\text{C}$ with no-core shell model (NCSM) returned a non-convergent result~\cite{Caurier:2002hb} (a new calculation with NCSM + continuum is on the way). We thus conclude that, a reliable methodology to study $\delta_\text{C}$ should at least be able to (directly or indirectly) take these two effects into account.

\section{\label{sec:strategy}Possible computational strategy}

Among the two criteria that we posted above for a reliable $\delta_\text{C}$ calculation, the requirement to sum over all excited-state contributions is equivalent to computing the inversion of the Hamiltonian matrix in the 
the nuclear Green's function $G(\zeta)$, which is computationally demanding. Fortunately, there are standard computational strategies to achieve this purpose, and here we provide one example based on the Lanczos algorithm~\cite{Lanczos:1950zz,Haydock_1974,Marchisio:2002jx}. 

We start by defining a properly-normalized starter state,
\begin{equation}
    |\phi_0\rangle\equiv \frac{M_{-1}^{(1)}|g;1,T_z\rangle}{\sqrt{\langle g;1,T_z|(M_{-1}^{(1)})^\dagger M_{-1}^{(1)}|g;1,T_z\rangle}}~.
\end{equation}
With this, the first term of $F_{T_z}(\zeta)$ can be written as
\begin{eqnarray}
   && \langle g;1,T_z|(M_{-1}^{(1)})^\dagger G(\zeta)M_{-1}^{(1)}|g;1,T_z\rangle\nonumber\\
   &=&\langle g;1,T_z|(M_{-1}^{(1)})^\dagger M_{-1}^{(1)}|g;1,T_z\rangle\langle\phi_0|G(\zeta)|\phi_0\rangle~.
\end{eqnarray}
Again, the coefficient $\langle g;1,T_z|(M_{-1}^{(1)})^\dagger M_{-1}^{(1)}|g;1,T_z\rangle$ only involves the ground state matrix element, while $\langle \phi_0|G(\zeta)|\phi_0\rangle$ is more complicated. To evaluate the latter, we construct a set of $n$ orthonormal Lanczos basis $\{|\phi_i\rangle\}_{i=0}^{n-1}$ through the following iteration:
\begin{equation}
    |w_{i+1}\rangle\equiv b_{i+1}|\phi_{i+1}\rangle\equiv H_0|\phi_i\rangle-a_i|\phi_i\rangle-b_i|\phi_{i-1}\rangle~,
\end{equation}
where 
\begin{equation}
    a_i\equiv \langle \phi_i|H_0|\phi_i\rangle~,~b_i\equiv\sqrt{\langle w_i|w_i\rangle}
\end{equation}
are the so-called Lanczos coefficients, with $b_0\equiv0$ and $|\phi_{-1}\rangle\equiv0$. The Hamiltonian $H_0$ is tridiagonalized under such basis, and the desired matrix element can be expressed as a continuous fraction involving the Lanczos coefficients,
\begin{equation}
    \langle \phi_0|G(\zeta)|\phi_0\rangle =g_0(\zeta)~,
\end{equation}
defined via the following recursion relation,
\begin{equation}
    g_i(\zeta)=\frac{1}{\zeta-a_i-b_{i+1}^2g_{i+1}(\zeta)}~,~i=0,1,...,n-2,\label{eq:recg}
\end{equation}
which terminates at $g_{n-1}(\zeta)=1/(\zeta-a_{n-1})$.
For completeness, we also provide the recursion relation of the first $\zeta$-derivative,
\begin{equation}
     g_i'(\zeta)=-g_i^2(\zeta)\left(1-b_{i+1}^2g_{i+1}'(\zeta)\right)~,~i=0,1,...,n-2,\label{eq:recgprime}
\end{equation}
with $g_{n-1}'(\zeta)=-g_{n-1}^2(\zeta)$. 

The procedure above determines $F_{T_z}(\zeta)$ completely in terms of two ground-state matrix elements $\langle g;1,T_z|(M_{-1}^{(1)})^\dagger M_{-1}^{(1)}|g;1,T_z\rangle$, $\langle g;1,T_z-1|M_{-1}^{(1)}|g;1,T_z\rangle$
and the Lanczos coefficients $\{a_i,b_i\}$, none of which requires a matrix inversion.
In particular, the expansion coefficients of our interest can be written as
\begin{eqnarray}
\alpha_{T_z}&=&\biggl\{\langle g;1,T_z|(M_{-1}^{(1)})^\dagger M_{-1}^{(1)}|g;1,T_z\rangle g_0(\zeta)\nonumber\\
&&-\frac{|\langle g;1,T_z-1|M_{-1}^{(1)}|g;1,T_z\rangle|^2}{\zeta-E_{g,1}}\biggr\}_{\zeta=E_{g,1}}\nonumber\\
\beta_{T_{z}}&=&\biggl\{\langle g;1,T_z|(M_{-1}^{(1)})^\dagger M_{-1}^{(1)}|g;1,T_z\rangle g_0'(\zeta)\nonumber\\
&&+\frac{|\langle g;1,T_z-1|M_{-1}^{(1)}|g;1,T_z\rangle|^2}{(\zeta-E_{g,1})^2}\biggr\}_{\zeta=E_{g,1}}~,
\end{eqnarray}
with $g_0(\zeta)$ and $g_0'(\zeta)$ entirely fixed by the Lanczos coefficients following Eqs.~\eqref{eq:recg}, \eqref{eq:recgprime}.
Within this formalism,  $\delta_\text{C}$ is tightly constrained by the experimental observables. 
To make a prediction of $\Delta M_{A,B}^{(1)}$, one needs to compute the two ground-state matrix elements and all the Lanczos coefficients. Once this is done, there is no more freedom left for $\delta_\text{C}$ which at this point can also be computed. 
The predicted values of $\Delta M_{A,B}^{(1)}$ can be compared  with the experiment, and the respective deviation and experimental uncertainty may be directly translated into the well-justified uncertainty estimate for $\delta_\text{C}$.

We stress that the strategy outlined above is, as such, model-independent.
Putting it into practice requires microscopic nuclear theory calculations of the ground-state matrix elements and the Lanczos coefficients, preferably with ab-initio methods. For light nuclei, methods such as Quantum Monte Carlo~\cite{Carlson:2014vla,Gandolfi:2020pbj} and NCSM~\cite{Barrett:2013nh} are powerful tools; for medium-size nuclei, coupled-cluster theory~\cite{Hagen:2013nca}, In-Medium Similarity Renormalization Group~\cite{Stroberg:2021guc} and nuclear lattice effective field theory~\cite{Lee:2004si,Borasoy:2006qn,Lee:2008fa,Lahde:2019npb} may be applicable. Notice that, for light nuclei ($A\sim 10$) some of our basic assumptions on ISB interactions (e.g. the isovector dominance) may not be as solid, but the definition of $F_{T_z}(\zeta)$ through Eq.\eqref{eq:FTz} is not affected by these assumptions and it can be computed nonetheless, which serves as important prototypes for future computations involving heavier nuclei. 
While the outlined strategy
is not based on a model, it uses several approximations. Such approximations include the identification of the ISB potential $V$ with the isovector monopole operator $\vec M^{(1)}$, the assumption of the uniform-sphere proton distribution, and the neglect of the isotensor part of $V$. The validity of these approximations should be subject of future studies.

\section{\label{sec:conclusion}Conclusions}

Despite the quoted high precision level of $|V_{ud}|_{0^+}$ in the literature, it has now become increasingly transparent that there could be hidden systematic uncertainties at the order $10^{-4}$ or larger, which were not reflected in the current error budget and are crucial for precision tests of SM at low energies. In particular, existing theory calculations of the ISB correction $\delta_\text{C}$ to the Fermi matrix element are model-dependent and, as we point out in this paper, may not be consistent with general constraints from isospin symmetry. We show that the ability of a nuclear theory approach to predict nuclear mass splittings does not imply the same predictive power for $\delta_\text{C}$: the former depend primarily on ground-state diagonal nuclear matrix elements, while the latter must involve excited states. On the other hand, the new ISB observables $\Delta M_{A,B}^{(1)}$ introduced in Ref.\cite{Seng:2022epj} are constructed from measurable electroweak nuclear radii, and probe the same nuclear matrix elements as $\delta_\text{C}$. Therefore, it is more natural to gauge the theory accuracy for $\delta_\text{C}$ using $\Delta M_{A,B}^{(1)}$, rather than the IMME coefficients. 

Existing ab-initio studies of $\delta_\text{C}$ consist mainly of direct computations of the full Fermi matrix element in the presence of ISB interactions. In this work we propose an alternative approach.  
Based on the isovector dominance of ISB interactions, we define the functions $F_{T_z}(\zeta)$ that involve matrix elements of isovector monopole operators and a single nuclear Green's function. We show that the coefficients of its expansion with respect to $\zeta$ around the ground state energy $E_{g,1}$ give simultaneously $\Delta M_{A,B}^{(1)}$ and $\delta_\text{C}$. With that, we recast the problem of an experimental-verifiable theory calculation of $\delta_\text{C}$ in terms of the study of the $\zeta$-dependence of $F_{T_z}(\zeta)$. Through a working example with conventional shell model, we conclude that the low-lying nuclear excitations are not the main contributors to a non-zero $\delta_\text{C}$, and speculate that the (1) summation over higher excitations and (2) continuum intermediate states may be important. The main difficulty to fulfill the first criteria is the inversion of a large Hamiltonian matrix in the nuclear Green's function $G(\zeta)$, which could be bypassed using mathematical techniques such as the Lanczos algorithm we described in Section~\ref{sec:strategy}. With this strategy, both $\delta_\text{C}$ and $\Delta M_{A,B}^{(1)}$ are uniquely determined from a set of ground-state nuclear matrix elements and Lanczos coefficients, and share the same level of the theoretical accuracy. 

Finally, we wish to point out a similarity between the new formalism for computing $\delta_\text{C}$  proposed in this work, and that for the nucleus-dependent radiative correction $\delta_\text{NS}$ introduced in Ref.\cite{Seng:2022cnq}. The latter depends on the generalized Compton tensor,
%\begin{equation}
$T^{\mu\nu}\sim \langle f|J_a^\mu G(\zeta) J_b^\nu|i\rangle$,
%\end{equation}
where $\{J_a^\mu,J_b^\nu\}$ are electroweak current operators and $G(\zeta)$ is the same nuclear Green's function that appears in this work. The only difference with $F_{T_z}$ is that the isovector monopole operators are replaced by current operators. Therefore, methodologies applicable for ab-initio calculations of $\delta_\text{NS}$ will also apply to $\delta_\text{C}$. This newly-identified similarity may help to promote simultaneous theory progress in the two quantities that are crucial for a precise extraction of $|V_{ud}|$, and further foster the potential of nuclear beta decay experiments for discovering or constraining  
new physics beyond the standard model. 

\begin{acknowledgments}

We thank Michael Gennari for many useful conversations, and express our gratitude to Alex Brown for introducing the NuShellX code to us.
The work of C.Y.S. is supported in
part by the U.S. Department of Energy (DOE), Office of Science, Office of Nuclear Physics, under the FRIB Theory Alliance award DE-SC0013617, and by the DOE grant DE-FG02-97ER41014. The work of M.G. is supported in part by EU Horizon 2020 research and innovation programme, STRONG-2020 project
under grant agreement No 824093, and by the Deutsche Forschungsgemeinschaft (DFG) under the grant agreement GO 2604/3-1. 

\end{acknowledgments}

\bibliography{deltaC_ref}

\end{document}